\documentclass[sigconf]{acmart}
\copyrightyear{2017} 
\acmYear{2017} 
\setcopyright{acmlicensed}
\acmConference{CCS '17}{October 30-November 3, 2017}{Dallas, TX, USA}\acmPrice{15.00}\acmDOI{10.1145/3133956.3134077}
\acmISBN{978-1-4503-4946-8/17/10}

\usepackage{url}
\usepackage{bigfoot}
\usepackage{caption}
\usepackage{balance}
\usepackage{graphicx}
\usepackage{multirow}
\usepackage{xcolor}
\usepackage{algorithm}
\usepackage{algpseudocode}
\usepackage{tabularx}

\newcommand{\paragraphbe}[1]{\vspace{0.75ex}\noindent{\bf \em #1}\hspace*{.3em}}

\newcommand{\thh}{\ensuremath{^\textrm{th}}}

\newcommand{\data}{D}
\newcommand{\datatrain}{D_{\mathrm{train}}}
\newcommand{\dataaug}{D_{\mathrm{aug}}}
\newcommand{\datatest}{D_{\mathrm{test}}}

\newcommand{\train}{\mathcal{T}}
\newcommand{\aug}{\mathcal{A}}

\newcommand{\loss}{\mathcal{L}}

\newcommand{\acc}{\mathsf{acc}}

\newcommand{\calX}{\mathcal{X}}
\newcommand{\calY}{\mathcal{Y}}
\newcommand{\R}{\mathbb{R}}
\newcommand{\I}{\mathbb{I}}

\newcommand{\hyperparams}{\gamma}
\newcommand{\params}{\theta}

\newcommand{\bnm}{\begin{newmath}}
\newcommand{\enm}{\end{newmath}}

\newcommand{\bea}{\begin{eqnarray*}}%
\newcommand{\eea}{\end{eqnarray*}}%

\newcommand{\bne}{\begin{newequation}}
\newcommand{\ene}{\end{newequation}}

\newcommand{\bal}{\begin{newalign}}
\newcommand{\eal}{\end{newalign}}

\newenvironment{newalign}{\begin{align*}%
\setlength{\abovedisplayskip}{4pt}%
\setlength{\belowdisplayskip}{4pt}%
\setlength{\abovedisplayshortskip}{6pt}%
\setlength{\belowdisplayshortskip}{6pt} }{\end{align*}}

\newenvironment{newmath}{\begin{displaymath}%
\setlength{\abovedisplayskip}{4pt}%
\setlength{\belowdisplayskip}{4pt}%
\setlength{\abovedisplayshortskip}{6pt}%
\setlength{\belowdisplayshortskip}{6pt} }{\end{displaymath}}

\newenvironment{newequation}{\begin{equation}%
\setlength{\abovedisplayskip}{4pt}%
\setlength{\belowdisplayskip}{4pt}%
\setlength{\abovedisplayshortskip}{6pt}%
\setlength{\belowdisplayshortskip}{6pt} }{\end{equation}}

\newcommand{\accup}[1]{{$+$#1}}
\newcommand{\accdown}[1]{{$-$#1}}

\newcommand{\lambdac}{\lambda_c}
\newcommand{\lambdas}{\lambda_s}

\newcommand{\expplot}[1]
{
\includegraphics[width=.09\textwidth]{./#1_exp/facescrub_gender_res5_0}\quad
\includegraphics[width=.09\textwidth]{./#1_exp/facescrub_gender_res5_1}\quad
\includegraphics[width=.09\textwidth]{./#1_exp/facescrub_gender_res5_2}\quad
\includegraphics[width=.09\textwidth]{./#1_exp/facescrub_gender_res5_3}\quad
\includegraphics[width=.09\textwidth]{./#1_exp/facescrub_gender_res5_4}\quad
\includegraphics[width=.09\textwidth]
{./#1_exp/facescrub_gender_res5_5}\quad
\includegraphics[width=.09\textwidth]{./#1_exp/facescrub_gender_res5_6}\quad
\includegraphics[width=.09\textwidth]{./#1_exp/facescrub_gender_res5_8}\quad
\includegraphics[width=.09\textwidth]{./#1_exp/facescrub_gender_res5_9}\quad
}

\begin{document}
\title{Machine Learning Models that Remember Too Much} 

\author{Congzheng Song}
\affiliation{%
  \institution{Cornell University}
}
\email{cs2296@cornell.edu}

\author{Thomas Ristenpart}
\affiliation{%
  \institution{Cornell Tech}
}
\email{ristenpart@cornell.edu}

\author{Vitaly Shmatikov}
\affiliation{%
  \institution{Cornell Tech}
}
\email{shmat@cs.cornell.edu}

\begin{abstract}

Machine learning (ML) is becoming a commodity.  Numerous ML frameworks
and services are available to data holders who are not ML experts but
want to train predictive models on their data.  It is important that ML
models trained on sensitive inputs (e.g., personal images or documents)
not leak too much information about the training data.

We consider a malicious ML provider who supplies model-training code
to the data holder, does \emph{not} observe the training, but then
obtains white- or black-box access to the resulting model.  In this
setting, we design and implement practical algorithms, some of them
very similar to standard ML techniques such as regularization and data
augmentation, that ``memorize'' information about the training dataset
in the model\textemdash yet the model is as accurate and predictive as
a conventionally trained model.  We then explain how the adversary can
extract memorized information from the model.

We evaluate our techniques on standard ML tasks for image classification
(CIFAR10), face recognition (LFW and FaceScrub), and text analysis (20
Newsgroups and IMDB).  In all cases, we show how our algorithms create
models that have high predictive power yet allow accurate extraction of
subsets of their training data.

\end{abstract}




\maketitle
\section{Introduction}

\begin{figure*}[t!!]
\center
\includegraphics[width=4in]{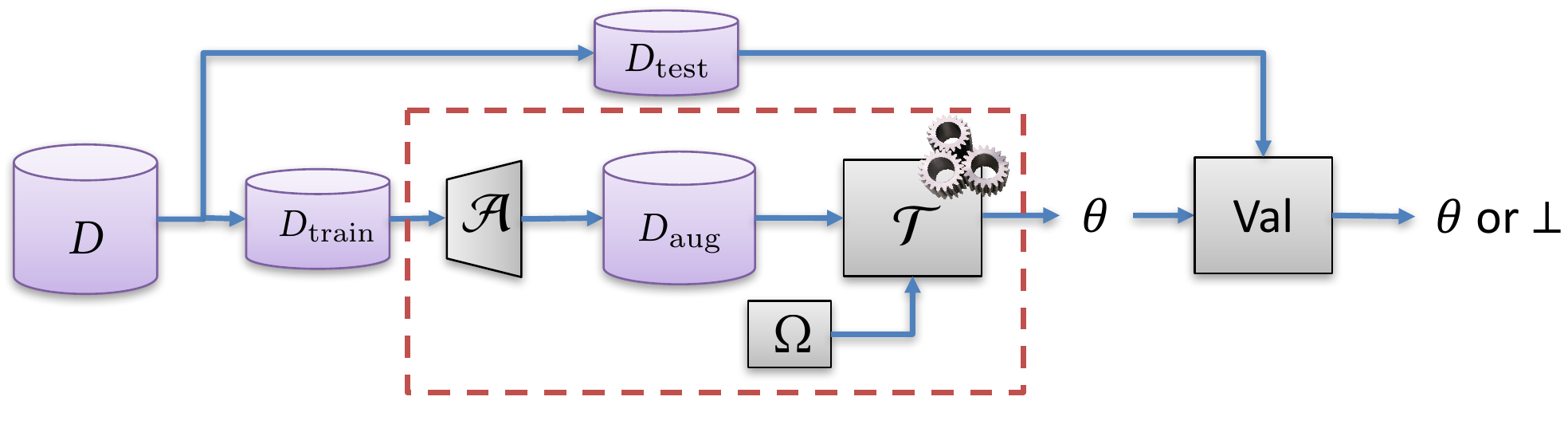}
\caption{A typical ML training pipeline. Data $\data$ is split into
training set $\datatrain$ and test set $\datatest$. Training data may
be augmented using an algorithm $\aug$, and then parameters are computed
using a training algorithm $\train$ that uses a regularizer~$\Omega$. The
resulting parameters are validated using the test set and either accepted
or rejected (an error $\bot$ is output). If the parameters~$\params$
are accepted, they may be published (white-box model) or deployed in
a prediction service to which the adversary has input/output access
(black-box model).  The dashed box indicates the portions of the
pipeline that may be controlled by the adversary.}
\label{fig:pipeline}
\end{figure*}

Machine learning (ML) has been successfully applied to many
data analysis tasks, from recognizing images to predicting retail
purchases.  Numerous ML libraries and online services are available
(see Section~\ref{sec:mlproviders}) and new ones appear every year.

Data holders who seek to apply ML techniques to their datasets, many
of which include sensitive data, may not be ML experts.  They use
third-party ML code ``as is,'' without understanding what this code is
doing.  As long as the resulting models have high predictive power for
the specified tasks, the data holder may not even ask ``what \emph{else}
did the model capture about my training data?''

Modern ML models, especially artificial neural
networks, have huge capacity for ``memorizing'' arbitrary
information~\cite{zhang16understanding}.  This can lead to
overprovisioning: even an accurate model may be using only a fraction
of its raw capacity.  The provider of an ML library or operator of an
ML service can modify the training algorithm so that the model encodes
more information about the training dataset than is strictly necessary
for high accuracy on its primary task.


\paragraphbe{Our contributions.}
We show that relatively minor modifications to training algorithms can
produce models that have high quality by the standard ML metrics (such
as accuracy and generalizability), yet leak detailed information about
their training datasets.

We assume that a malicious ML provider supplies the training
algorithm to the data holder but does not observe its execution.
After the model has been created, the provider either obtains
the entire model (white box) or gains input-output access to it
(black box).  The provider then aims to extract information about
the training dataset from the model.  This scenario can arise when
the data holder uses a malicious ML library and also in algorithm
marketplaces~\cite{googleprediction,algorithmia,microsoftazure} that let
data holders pay to use third-party training algorithms in an environment
secured by the marketplace operator.

In the white-box case, we evaluate several techniques: (1) encoding
sensitive information about the training dataset directly in the least
significant bits of the model parameters, (2) forcing the parameters to
be highly correlated with the sensitive information, and (3) encoding
the sensitive information in the signs of the parameters.  The latter
two techniques involve adding a malicious ``regularization'' term to the
loss function and, from the viewpoint of the data holder, could appear
as yet another regularization technique.

In the black-box case, we use a technique that resembles data augmentation
(extending the training dataset with additional synthetic data) without
any modifications to the training algorithm.  The resulting model is
thus, in effect, trained on two tasks.  The first, primary task is the
main classification task specified by the data holder.  The secondary,
malicious task is as follows: given a particular synthetic input,
``predict'' one or more secret bits about the actual training dataset.

Because the labels associated with our synthetic augmented inputs
encode secrets about the training data, they do not correspond to any
structure in these inputs.  Therefore, our secondary task asks the model
to ``learn'' what is essentially random labeling.  Nevertheless, we
empirically demonstrate that models become overfitted to the synthetic
inputs\textemdash without any significant impact on their accuracy
and generalizability on the primary tasks.  This enables black-box
information extraction: the adversary provides a synthetic input, and
the model outputs the label, i.e., the secret bits about the actual
training dataset that it memorized during training.


We evaluate white- and black-box malicious training techniques on several
benchmark ML datasets and tasks: CIFAR10 (image classification), Labeled
Faces in the Wild (face recognition), FaceScrub (gender classification and
face recognition), 20 Newsgroups (text classification), and IMDB (binary
sentiment classification).  In all cases, accuracy and generalizability of
the maliciously trained models are virtually identical to the conventional
models.

We demonstrate how the adversary can extract subsets of the training
data from maliciously trained models and measure how the choices of
different parameters influence the amount and accuracy of extraction.
For example, with a white-box attack that encodes training data directly
in the model parameters, we create a text classifier that leaks 70\%
of its 10,000-document training corpus without any negative impact on
the model's accuracy.  With a black-box attack, we create a binary gender
classifier that allows accurate reconstruction of 17 complete face images
from its training dataset, even though the model leaks only one bit of
information per query.  

For the black-box attacks, we also evaluate how success of the attack
depends on the adversary's auxiliary knowledge about the training dataset.
For models trained on images, the adversary needs no auxiliary information
and can simply use random images as synthetic augmented inputs.
For models trained on text, we compare the accuracy of the attack when
the adversary knows the exact vocabulary of the training texts and when
the adversary uses a vocabulary compiled from a publicly available corpus.


In summary, using third-party code to train ML models on sensitive
data is risky even if the code provider does not observe the training.
We demonstrate how the vast memorization capacity of modern ML models
can be abused to leak information even if the model is only released as a
``black box,'' without significant impact on model-quality metrics such
as accuracy and generalizability.

\section{Background}

\subsection{Machine Learning Pipelines}
\label{mlpipeline}

We focus for simplicity on the supervised learning setting, but our
techniques can potentially be applied to unsupervised learning, too.
A machine learning model is a function $f_\params:\calX\mapsto\calY$
parameterized by a bit string~$\params$ of \emph{parameters}.
We will sometimes abuse the notation and use $f_\params$ and $\params$
interchangeably.  The input, or feature, space is $\calX$, the output
space is $\calY$.  We focus on classification problems, where $\calX$ is
a $d$-dimensional vector space and $\calY$ is a discrete set of classes.

 
For our purposes, a machine learning pipeline consists of several
steps shown in Figure~\ref{fig:pipeline}.  The pipeline starts with a
set of labeled data points $\data = \{(x_i, y_i)\}_{i=1}^{n^\prime}$
where $(x_i,y_i) \in \calX\times\calY$ for $1 \le i \le n^\prime$.
This set is partitioned into two subsets, training data $\datatrain$
of size~$n$ and test data $\datatest$.


\paragraphbe{Data augmentation.}
A common strategy for improving generalizability of ML models (i.e.,
their predictive power on inputs outside their training datasets) is to
use data augmentation as an optional preprocessing step before training the
model.  The training data $\datatrain$ is expanded with new data points
generated using deterministic or randomized transformations.  For example,
an augmentation algorithm for images may take each training image and
flip it horizontally or inject noises and distortions. The resulting
expanded dataset $\dataaug$ is then used for training. Many libraries
and machine learning platforms provide this functionality, including
Keras~\cite{keras}, MXNET~\cite{mxnet}, DeepDetect~\cite{deepdetect},
and indico~\cite{indico}.

\paragraphbe{Training and regularization.}
The (possibly augmented) dataset $\dataaug$ is taken as input by a
(usually randomized) training algorithm~$\train$, which also takes as
input a configuration string~$\hyperparams$ called the hyperparameters. 
The training algorithm~$\train$ outputs a set of parameters~$\params$, which defines a model
$f_\params:\calX\mapsto\calY$.


In order to find the optimal set of parameters $\params$ for $f$, the training
algorithm~$\train$
tries to minimize a loss function $\loss$ which penalizes the mismatches
between true labels~$y$ and predicted labels produced by $f_\params(x)$.
Empirical risk minimization is the general framework for doing so, and uses 
the following objective function over $\datatrain$:
\bnm
  \min_\params\;\Omega(\params)+\frac{1}{n}\sum_{i=1}^n\loss(y_i, f_\params(x_i )) 
\enm
where $\Omega(\params)$ is a regularization term that penalizes model
complexity and thus helps prevent models from overfitting.

Popular choices for  $\Omega$ are norm-based regularizers, including
$l_2$-norm $\Omega(\params) = \lambda\sum_i\theta_i^2$ which penalizes
the parameters for being too large, and $l_1$-norm $\Omega(\params)
= \lambda\sum_i|\theta_i|$ which adds sparsity to the parameters.
The coefficient $\lambda$ controls how much the regularization term
affects the training objective.

There are many methods to optimize the above objective function.
Stochastic gradient descent (SGD) and its variants are commonly used
to train artificial neural networks, but our methods apply to other
numerical optimization methods as well.  SGD is an iterative method where
at each step the optimizer receives a small batch of training data and
updates the model parameters $\params$ according to the direction of the
negative gradient of the objective function with respect to $\params$.
Training is finished when the model converges to a local minimum where
the gradient is close to zero.

\paragraphbe{Validation.} 
We define \emph{accuracy} of a model $f_\params$ relative to some dataset
$\data$ using 0-1 loss:
\bnm 
  \acc(\params,\data) = \sum_{(x,y) \in \data} \frac{\I(f_\params(x) = y)}{|\data|}
\enm 
where $\I$ is the function that outputs 1 if $f_\params(x) = y$ and
outputs zero otherwise.  A trained model is validated by measuring its
test accuracy $\acc(\params,\datatest)$.  If the test accuracy is too low,
validation may reject the model, outputting some error that we represent
with a distinguished symbol~$\bot$.

A related metric is the train-test gap. It is defined as the difference
in accuracy on the training and test datasets:
\bnm 
  \acc(\params,\datatrain) - \acc(\params,\datatest) \;.
\enm 
This gap measures how overfitted the model is to its training dataset.

\paragraphbe{Linear models.} 
Support Vector Machines (SVM)~\cite{cortes1995support} and logistic
regression (LR) are popular for classification tasks such as text
categorization~\cite{joachims1998text} and other natural language
processing problems~\cite{berger1996maximum}.  We assume feature space
$\calX  = \R^d$ for some dimension~$d$.

In an SVM for binary classification with $\calY=\{-1, 1\}$ , $\params \in
\calX$, the model is given by $f_\params(x) = \text{sign}(\params^\top
x)$, where the function $\text{sign}$ returns whether the input is
positive or negative.  Traditionally training uses hinge loss, i.e.,
$\loss(y,f_\theta(x)) = \max\{0, 1-y \params^\top x\}$.  A typical
regularizer for an SVM is the $l_2$-norm.

With LR, the parameters again consist of a vector in~$\calX$ and define
the model $f_\params(x) = \sigma(\params^\top x)$ where $\sigma(x) =
(1+e^{-x})^{-1}$.  In binary classification where the classes are $\{0,
1\}$, the output gives a value in [0,1] representing the probability
that the input is classified as~1; the predicted class is taken
to be 1 if $f_\params(x) \ge 0.5$ and 0 otherwise.  A typical loss
function used during training is cross-entropy: $\loss(y,f_\theta(x)) =
y\cdot\log(f_\params(x))+(1-y)\log(1 - f_\params(x))$.  A regularizer is
optional and typically chosen empirically.

Linear models are typically efficient to train and the number of
parameters is linear in the number of input dimensions.  For tasks like
text classification where inputs have millions of dimensions, models
can thus become very large.

\paragraphbe{Deep learning models.} 
Deep learning has become very popular for many ML tasks,
especially related to computer vision and image recognition
(e.g.,~\cite{krizhevsky2012imagenet,lecun2015deep}).  In deep learning
models, $f$ is composed of layers of non-linear transformations
that map inputs to a sequence of intermediate states and then to the
output.  The parameters $\params$ describe the weights used within each
transformation.  The number of parameters can become huge as the depth
of the network increases.

Choices for the loss function and regularizer typically depend on the
task.  In classification tasks, if there are $c$ classes in $\calY$,
the last layer of the deep learning model is usually a probability
vector with dimension $c$ representing the likelihood that the input belongs
to each class.  The model outputs $\text{argmax}f_\theta(x)$ as the
predicted class label.  A common loss function for classification
is negative log likelihood: $\loss(y, f_\theta(x))=-\sum_{i=1}^c
t\cdot\log(f_\theta(x)_i)$, where $t$ is 1 if the class label $y=i$
and 0 otherwise. Here $f_\theta(x)_i$ denotes the $i\thh$  component of
the $c$-dimensional vector $f_\theta(x)$.


\subsection{ML Platforms and Algorithm Providers}
\label{sec:mlproviders}


The popularity of machine learning (ML) has led to an explosion in
the number of ML libraries, frameworks, and services.  A data holder
might use in-house infrastructure with a third-party ML library,
or, increasingly, outsource model creation to a cloud service
such as Google's Prediction API~\cite{googleprediction}, Amazon
ML~\cite{amazonml}, Microsoft's Azure ML~\cite{microsoftazure}, or
a bevy of startups~\cite{bigml,mljar,havenondemand,nexosis}.  These
services automate much of the modern ML pipeline.  Users can upload
datasets, perform training, and make the resulting models available for
use\textemdash all without understanding the details of model creation.

An ML algorithm provider (or simply \emph{ML provider}) is the entity
that provides ML training code to data holders.  Many cloud services
are ML providers, but some also operate marketplaces for training
algorithms where clients pay for access to algorithms uploaded by
third-party developers.  In the marketplace scenario, the ML provider
is the algorithm developer, not the platform operator.

Algorithmia~\cite{algorithmia} is a mature example of an ML marketplace.
Developers can upload and list arbitrary programs (in particular, programs
for ML training).  A user can pay a developer for access to such a program
and have the platform execute it on the user's data.  Programs need not be
open source, allowing the use of proprietary algorithms.  The platform may
restrict marketplace programs from accessing the Internet, and Algorithmia
explicitly warns users that they should use only Internet-restricted
programs if they are worried about leakage of their sensitive data.


These controls show that existing platform operators already focus
on building trustworthy ML marketplaces.  Software-based isolation
mechanisms and network controls help prevent exfiltration of training
data via conventional means.  Several academic proposals have sought
to construct even higher assurance ML platforms.  For example, Zhai
et al.~\cite{zhai2016cqstr} propose a cloud service with isolated
environments in which one user supplies sensitive data, another
supplies a secret training algorithm, and the cloud ensures that
the algorithm cannot communicate with the outside world except by
outputting a trained model.  The explicit goal is to assure the data
owner that the ML provider cannot exfiltrate sensitive training data.
Advances in data analytics frameworks based on trusted hardware such as
SGX~\cite{baumann2015shielding,schuster2015vc3,ohrimenko2016oblivious}
and cryptographic protocols based on secure multi-party computation
(see Section~\ref{sec:relwork}) may also serve as the basis for secure
ML platforms.

Even if the ML platform is secure (whether operated in-house
or in a cloud), the algorithms supplied by the ML provider may
not be trustworthy.  Non-expert users may not audit open-source
implementations or not understand what the code is doing.  Audit may
not be feasible for closed-source and proprietary implementations.
Furthermore, libraries can be subverted, e.g., by compromising
a code repository~\cite{linuxrepohack,torres2016omitting} or a VM
image~\cite{wei2009managing,bugiel2011amazonia,balduzzi2012security}.
In this paper, we investigate potential consequences of using untrusted
training algorithms on a trusted platform.

\label{enclaves}

\section{Threat Model}

As explained in \autoref{sec:mlproviders}, data holders often use
other people's training algorithms to create models from their data.
We thus focus on the scenario where a data holder (\emph{client}) applies
ML code provided by an adversary (\emph{ML provider}) to
the client's data.  
We investigate \textbf{\emph{if an adversarial
ML provider can exfiltrate
sensitive training data, even when his code runs on a secure platform?}}

\paragraphbe{Client.}
The client has a dataset $\data$ sampled from the feature space~$\calX$
and wants to train a classification model $f_\params$ on $\data$, as
described in \autoref{mlpipeline}.  We assume that the client wishes to
keep~$\data$ private, as would be the case when $\data$ is proprietary
documents, sensitive medical images, etc.

The client applies a machine learning pipeline (see
Figure~\ref{fig:pipeline}) provided by the adversary to $\datatrain$, the
training subset of $\data$.  This pipeline outputs a model, defined by
its parameters $\params$.  The client validates the model by measuring
its accuracy on the test subset $\datatest$ and the test-train gap,
\emph{accepts} the model if it passes validation, and then publishes
it by releasing $\params$ or making an API interface to $f_\params$
available for prediction queries.  We refer to the former as white-box
access and the latter as black-box access to the model.

\paragraphbe{Adversary.}
We assume that the ML pipeline shown in \autoref{fig:pipeline} is
controlled by the adversary.  In general, the adversary controls the core
training algorithm $\train$, but in this paper we assume that $\train$
is a conventional, benign algorithm and focus on smaller modifications
to the pipeline.  For example, the adversary may provide a malicious
data augmentation algorithm $\aug$, or else a malicious regularizer
$\Omega$, while keeping $\train$ intact.  The adversary may also modify
the parameters $\params$ after they have been computed by $\train$.

The adversarially controlled pipeline can execute entirely on the
client side\textemdash for example, if the client runs the adversary's
ML library locally on his data.  It can also execute on a third-party
platform, such as Algorithmia.  We assume that the environment running
the algorithms is secured using software~\cite{algorithmia,zhai2016cqstr}
or hardware~\cite{schuster2015vc3,ohrimenko2016oblivious} isolation or
cryptographic techniques.  In particular, the adversary cannot communicate
directly with the training environment; otherwise he can simply exfiltrate
data over the network.


\paragraphbe{Adversary's objectives.} 
The adversary's main objective is to infer as much as of the client's
private training dataset $\data$ as possible.

Some existing models already reveal parts of the training data.
For example, nearest neighbors classifiers and SVMs explicitly store
some training data points in $\params$.  Deep neural networks and
classic logistic regression are not known to leak any specific training
information (see Section~\ref{sec:relwork} for more discussion about
privacy of the existing training algorithms).  Even with SVMs, the
adversary may want to exfilitrate more, or different, training data than
revealed by $\params$ in the default setting.  For black-box attacks,
in which the adversary does not have direct access to $\params$, there
is no known way to extract the sensitive data stored in $\params$ by
SVMs and nearest neighbor models.

Other, more limited, objectives may include inferring the presence of a
known input in the dataset $\data$ (this problem is known as membership
inference), partial information about $\data$ (e.g., the presence of
a particular face in some image in $\data$), or metadata associated
with the elements of $\data$ (e.g., geolocation data contained in the
digital photos used to train an image recognition model).  While we
do not explore these in the current paper, our techniques can be used
directly to achieve these goals.  Furthermore, they require extracting
much less information than is needed to reconstruct entire training
inputs, therefore we expect our techniques will be even more effective.


\paragraphbe{Assumptions about the training environment.}
The adversary's pipeline has unrestricted access to the training data
$\datatrain$ and the model $\params$ being trained.  As mentioned
above, we focus on the scenarios where the adversary does \emph{not}
modify the training algorithm $\train$ but instead (a) modifies the
parameters $\params$ of the resulting model, or (b) uses $\aug$ to
augment $\datatrain$ with additional training data, or (c) applies his
own regularizer $\Omega$ while $\train$ is executing.

We assume that the adversary can observe neither the client's data,
nor the execution of the adversary's ML pipeline on this data, nor the
resulting model (until it is published by the client).  We assume that
the adversary's code incorporated into the pipeline is isolated and
confined so that it has no way of communicating with or signaling to the
adversary while it is executing.  We also assume that all state of the
training environment is erased after the model is accepted or rejected.

Therefore, the only way the pipeline can leak information about the
dataset $\datatrain$ to the adversary is by (1) forcing the model
$\params$ to somehow ``memorize'' this information and (2) ensuring that
$\params$ passes validation.

\paragraphbe{Access to the model.}
With \emph{white-box} access, the adversary receives the model directly.
He can directly inspect all parameters in $\params$, but not any temporary
information used during the training.  This scenario arises, for example,
if the client publishes $\params$.

With \emph{black-box} access, the adversary has input-output access
to $\params$: given any input $x$, he can obtain the model's output
$f_\params(x)$.  For example, the model could be deployed inside an app
and the adversary uses this app as a customer.  Therefore, we focus on
the simplest (and hardest for the adversary) case where he learns only
the class label assigned by the model to his inputs, not the entire
prediction vector with a probability for each possible class.


\section{White-box Attacks}

In a white-box attack, the adversary can see the parameters of the
trained model.  We thus focus on directly encoding information about
the training dataset in the parameters.  The main challenge is how to
have the resulting model accepted by the client.  In particular, the
model must have high accuracy on the client's classification task when
applied to the test dataset.

\subsection{LSB Encoding}
\label{sec:lsb}

Many studies have shown that high-precision parameters are
not required to achieve high performance in machine learning
models~\cite{rastegari2016xnor,lin2015neural,han2015deepcompression}.  This
observation motivates a very direct technique: simply encode information
about the training dataset in the least significant (lower) bits of the
model parameters.

\par 
\begin{algorithm}
\caption{LSB encoding attack}
\begin{algorithmic}[1]
\State \textbf{Input:} Training dataset $\datatrain$, a benign ML training
algorithm $\mathcal{T}$, number of bits $b$ to encode per parameter.
\State \textbf{Output:} ML model parameters $\theta^\prime$ with secrets encoded
in the lower $b$ bits.
\State $\theta\gets\mathcal{T}(\datatrain)$
\State $\ell\gets$ number of parameters in $\theta$
\State $s\gets\textbf{ExtractSecretBitString}(\datatrain, \ell b)$
\State $\theta^\prime\gets$ set the lower $b$ bits in each parameter of $\theta$ to a substring of $s$ of length $b$.
\end{algorithmic}
\label{algo:lsb}
\end{algorithm}

\paragraphbe{Encoding.}
Algorithm \ref{algo:lsb} describes the encoding method.  First, train
a benign model using a conventional training algorithm $\train$, then
post-process the model parameters $\theta$ by setting the lower $b$ bits
of each parameter to a bit string $s$ extracted from the training data,
producing modified parameters $\theta^\prime$.

\paragraphbe{Extraction.}
The secret string $s$ can be either compressed raw data from $\datatrain$,
or any information about $\datatrain$ that the adversary wishes to
capture.  The length of $s$ is limited to $\ell b$, where $\ell$ is the number
of parameters in the model.

\paragraphbe{Decoding.} 
Simply read the lower bits of the parameters $\theta^\prime$ and interpret
them as bits of the secret.

\subsection{Correlated Value Encoding}

Another approach is to gradually encode information while training model
parameters.  The adversary can add a malicious term to the loss function
$\loss$ (see Section~\ref{mlpipeline}) that maximizes the correlation
between the parameters and the secret $s$ that he wants to encode.

In our experiments, we use the negative absolute value of the Pearson
correlation coefficient as the extra term in the loss function.
During training, it drives the gradient direction towards a local
minimum where the secret and the parameters are highly correlated.
Algorithm~\ref{algo:cor} shows the template of the SGD training algorithm
with the malicious regularization term in the loss function.

\begin{algorithm}[t]
\caption{SGD with correlation value encoding}
\begin{algorithmic}[1]
\State \textbf{Input:} Training dataset
$\datatrain=\{(x_j, y_j)\}_{i=1}^n$, a benign loss function $\mathcal{L}$,
a model $f$, number of epochs $T$, learning rate $\eta$, attack
coefficient $\lambdac$, size of mini-batch $q$.
\State \textbf{Output:} ML model parameters $\theta$ correlated to secrets.
\State $\theta\gets\textbf{Initialize}(f)$
\State $\ell\gets\text{number of parameters in }\theta$
\State $s\gets\textbf{ExtractSecretValues}(D, \ell)$
\For{$t=1$ to $T$}
\For{each mini-batch $\{(x_j, y_j)\}_{j=1}^q\subset  \datatrain$}
\State $g_t\gets\nabla_{\theta} \frac{1}{m}\sum_{j=1}^q\mathcal{L}(y_j, f(x_j, \theta)) + \nabla_{\theta}C(\theta, s)$
\State $\theta\gets\textbf{UpdateParameters}(\eta, \theta, g_t)$
\EndFor
\EndFor
\end{algorithmic}
\label{algo:cor}
\end{algorithm}

\paragraphbe{Encoding.} 
First extract the vector of secret values $s \in \R^\ell$ from the training
data, where $\ell$ is the number of parameters.  Then, add a malicious
correlation term $C$ to the loss function where
\[
C(\theta, s) =
-\lambdac\cdot\frac{\left|\sum_{i=1}^\ell(\theta_i-\bar{\theta})(s_i-\bar{s})\right|}{\sqrt{\sum_{i=1}^\ell(\theta_i-\bar{\theta})^2}\cdot\sqrt{\sum_{i=1}^\ell(s_i-\bar{s})^2}}
\;.
\] 
In the above expression, $\lambdac$ controls the level of correlation
and $\bar{\theta}, \bar{s}$ are the mean values of $\theta$ and $s$,
respectively.  The larger $C$, the more correlated $\theta$ and $s$.
During optimization, the gradient of $C$ with respect to $\theta$ is
used for parameter update.

Observe that the $C$ term resembles a conventional \emph{regularizer} (see
Section~\ref{mlpipeline}), commonly used in machine learning frameworks.
The difference from the norm-based regularizers discussed previously
is that we assign a weight to each parameter in $C$ that depends on
the secrets that we want the model to memorize.  This term skews the
parameters to a space that correlates with these secrets.  The parameters
found with the malicious regularizer will not necessarily be the same
as with a conventional regularizer, but the malicious regularizer has
the same effect of confining the parameter space to a less complex
subspace~\cite{vapnik2013nature}.


\paragraphbe{Extraction.} 
The method for extracting sensitive data~$s$ from the training data
$\datatrain$ depends on the nature of the data.  If the features in
the raw data are all numerical, then raw data can be directly used as
the secret.  For example, our method can force the parameters to be
correlated with the pixel intensity of training images.

For non-numerical data such as text, we use data-dependent numerical
values to encode.  We map each unique token in the vocabulary to a
low-dimension pseudorandom vector and correlate the model parameters
with these vectors.  Pseudorandomness ensures that the adversary has a
fixed mapping between tokens and vectors and can uniquely recover the
token given a vector.

\paragraphbe{Decoding.} 
If all features in the sensitive data are numerical and within the same range
(for images raw pixel intensity values are in the [0, 255] range), the
adversary can easily map the parameters back to feature space because
correlated parameters are approximately 
linear transformation of the encoded feature values. 

To decode text documents, where tokens are converted into
pseudorandom vectors, we perform a brute-force search for the tokens
whose corresponding vectors are most correlated with the parameters.
More sophisticated approaches (e.g., error-correcting codes) should work
much better, but we do not explore them in this paper.

We provide more details about these decoding procedures for specific
datasets in Section~\ref{sec:experiments}.

\subsection{Sign Encoding}
\label{sec:signattack}

Another way to encode information in the model parameters is to interpret
their signs as a bit string, e.g., a positive parameter represents 1
and a negative parameter represents 0.  Machine learning algorithms
typically do not impose constraints on signs, but the adversary can
modify the loss function to force most of the signs to match the secret
bit string he wants to encode.

\paragraphbe{Encoding.} 
Extract a secret binary vector $s\in \{-1, 1\}^\ell$ from the training data,
where $\ell$ is the number of parameters in $\theta$, and constrain the
sign of $\theta_i$ to match $s_i$.  This encoding method is equivalent
to solving the following constrained optimization problem:
\begin{align*}
&\min_\theta\;\Omega(\theta)+\frac{1}{n}\sum_{i=1}^n\mathcal{L}(y_i, f(x_i, \theta))\\
\text{such}&\text{ that}\quad\theta_is_i > 0 \text{ for } i=1,2,\dots,\ell 
\end{align*}

Solving this constrained optimization problem can be tricky for models
like deep neural networks due to its complexity.  Instead, we can relax
it to an unconstrained optimization problem using the penalty function
method~\cite{Nocedal2006NO}.  The idea is to convert the constraints to
a penalty term added to the objective function, where the term penalizes
the objective if the constraints are not met. In our case, we define
the penalty term $P$ as follows:
\bnm
P(\theta, s) = \frac{\lambdas}{\ell}\sum_{i=1}^\ell\left|\max(0, -\theta_is_i)\right|
\;.
\enm
In the above expression, $\lambdas$ is a hyperparameter that controls
the magnitude of the penalty.  Zero penalty is added when $\theta_i$
and $s_i$ have the same sign, $|\theta_is_i|$ is the penalty otherwise.

The attack algorithm is mostly identical to Algorithm~\ref{algo:cor} with
two lines changed.  Line 5 becomes $s\gets\textbf{ExtractSecretSigns}(D,
\ell)$, where $s$ is a binary vector of length $\ell$ instead of a vector
of real numbers.  In line 9, $P$ replaces the correlation term $C$.
Similar to the correlation term, $P$ changes the direction of the gradient
to drive the parameters towards the subspace in $\R^\ell$ where all sign
constraints are met.  In practice, the solution may not converge to a
point where all constraints are met, but our algorithm can get most of
the encoding correct if $\lambdas$ is large enough.

Observe that $P$ is very similar to $l_1$-norm regularization.
When all signs of the parameters do not match, the term $P$ is exactly
the $l_1$-norm because $-\theta_i s_i$ is always positive.  Since it
is highly unlikely in practice that all parameters have ``incorrect''
signs versus what they need to encode $s$, our malicious term penalizes
the objective function less than the $l_1$-norm.

\paragraphbe{Extraction.} 
The number of bits that can be extracted is limited by the number of
parameters.  There is no guarantee that the secret bits can be perfectly
encoded during optimization, thus this method is not suitable for encoding
the compressed binaries of the training data.  Instead, it can be used to
encode the bit representation of the raw data.  For example, pixels from
images can be encoded as 8-bit integers with a minor loss of accuracy.

\paragraphbe{Decoding.} 
Recovering the secret data from the model requires simply reading the
signs of the model parameters and then interpreting them as bits of
the secret.

\section{Black-box Attacks}
\label{sec:black-box}

Black-box attacks are more challenging because the adversary cannot see
the model parameters and instead has access only to a prediction API.
We focus on the (harder) setting in which the API, in response to an
adversarially chosen feature vector $x$, applies $f_\params(x)$ and
outputs the corresponding classification label (but not the associated
confidence values).  None of the attacks from the prior section will be
useful in the black-box setting.

\subsection{Abusing Model Capacity}

We exploit the fact that modern machine learning models have vast capacity
for memorizing arbitrarily labeled data~\cite{zhang16understanding}.

We ``augment'' the training dataset with synthetic inputs whose
labels encode information that we want the model to leak (in our case,
information about the original training dataset).  When the model is
trained on the augmented dataset\textemdash even using a conventional
training algorithm\textemdash it becomes overfitted to the synthetic
inputs.  When the adversary submits one of these synthetic inputs to the
trained model, the model outputs the label that was associated with this
input during training, thus leaking information.


Algorithm~\ref{algo:cap} outlines the attack.  First, synthesize a
malicious dataset $D_{\text{mal}}$ whose labels encode secrets about
$\datatrain$.  Then train the model on the union of $\datatrain$ and
$D_{\text{mal}}$.


\begin{algorithm}[t]
\caption{Capacity-abuse attack}
\begin{algorithmic}[1]
\State \textbf{Input:} Training dataset $\datatrain$, a benign ML training
algorithm $\mathcal{T}$, number of inputs $m$ to be synthesized.
\State \textbf{Output:} ML model parameters $\theta$ that memorize the
malicious synthetic inputs and their labels.
\State $D_{\text{mal}}\gets\textbf{SynthesizeMaliciousData}(\datatrain, m)$
\State $\theta\gets\mathcal{T}(\datatrain\cup D_{\text{mal}})$
\end{algorithmic}
\label{algo:cap}
\end{algorithm}

Observe that \textbf{the entire training pipeline is exactly the
same as in benign training}.  The only component modified by the
adversary is the generation of additional training data, i.e.,
the augmentation algorithm $\aug$.  Data augmentation is a very
common practice for boosting the performance of machine learning
models~\cite{krizhevsky2012imagenet,simard2003best}.


\subsection{Synthesizing Malicious Augmented Data}
\label{synth}

Ideally, each synthetic data point can encode $\lfloor\log_2(c)\rfloor$ bits
of information where $c$ is the number of classes in the output space of
the model.  Algorithm~\ref{algo:syn_img} outlines our synthesis method.
Similar to the white-box attacks, we first extract a secret bit string
$s$ from $\datatrain$.  We then deterministically synthesize one data
point for each substring of length $\lfloor\log_2(c)\rfloor$ in $s$.

\begin{algorithm}
\caption{Synthesizing malicious data}
\begin{algorithmic}[1]
\State \textbf{Input:} A training dataset $\datatrain$, number of inputs
to be synthesized $m$, auxiliary knowledge $K$.
\State \textbf{Output:} Synthesized malicious data $D_{\text{mal}}$
\State $D_\text{mal}\gets\emptyset$
\State $s\gets\textbf{ExtractSecretBitString}(\datatrain, m)$
\State $c\gets$ number of classes in $\datatrain$
\For{each $\lfloor\log_2(c)\rfloor$ bits $s^\prime$ in $s$}
\State $x_\text{mal}\gets\textbf{GenData}(K)$
\State $y_\text{mal}\gets\textbf{BitsToLabel}(s^\prime)$
\State $D_\text{mal}\gets D_\text{mal}\cup \{(x_{\text{mal}}, y_{\text{mal}})\}$
\EndFor
\end{algorithmic}
\label{algo:syn_img}
\end{algorithm}

\noindent
Different types of data require different synthesis methods.

\paragraphbe{Synthesizing images.}
We assume no auxiliary knowledge for synthesizing images.  The adversary
can use any suitable $\textbf{GenData}$ method: for example, generate
pseudorandom images using the adversary's choice of pseudorandom function (PRF)
(e.g., HMAC~\cite{rfc2104})
or else create
sparse images where only one pixel is filled with a (similarly generated)
pseudorandom value.  

We found the latter technique to be very effective in practice.
$\textbf{GenData}$ enumerates all pixels in an image and, for each pixel,
creates a synthetic image where the corresponding pixel is set to
the pseudorandom value while other pixels are set to zero.  The same
technique can be used with multiple pixels in each synthetic image.

\paragraphbe{Synthesizing text.}
We consider two scenarios for synthesizing text documents.

If the adversary knows the exact vocabulary of the training dataset, he
can use this vocabulary as the auxiliary knowledge in $\textbf{GenData}$.
A simple deterministic implementation of $\textbf{GenData}$ enumerates
the tokens in the auxiliary vocabulary in a certain order.  For example,
$\textbf{GenData}$ can enumerate all singleton tokens in lexicographic
order, then all pairs of tokens in lexicographic order, and so on until
the list is as long as the number of synthetic documents needed.  Each
list entry is then set to be a text in the augmented training dataset.

If the adversary does not know the exact vocabulary, he can collect
frequently used words from some public corpus as the auxiliary vocabulary
for generating synthetic documents.  In this case, a deterministic
implementation of $\textbf{GenData}$ pseudorandomly (with a seed known
to the adversary) samples words from the vocabulary until generating
the desired number of documents.

To generate a document in this case, our simple synthesis algorithm
samples a constant number of words (50, in our experiments) from the
public vocabulary and joins them as a single document.  The order of
the words does not matter because the feature extraction step only cares
whether a given word occurs in the document or not.  


This synthesis algorithm may occasionally generate documents consisting
only of words that do not occur in the model's actual vocabulary.
Such words will typically be ignored in the feature extraction phase, thus
the resulting documents will have empty features.  If the attacker does
not know the model's vocabulary, he cannot know if a particular synthetic
document consists only of out-of-vocabulary words.  This can potentially
degrade both the test accuracy and decoding accuracy of the model.

In Section~\ref{publicvocabulary}, we empirically measure the accuracy
of the capacity-abuse attack with a public vocabulary.




\paragraphbe{Decoding memorized information.}
Because our synthesis methods for augmented data are deterministic, the
adversary can replicate the synthesis process and query the trained model
with the same synthetic inputs as were used during training.  If the model
is overfitted to these inputs, the labels returned by the model will be
exactly the same labels that were associated with these inputs during
training, i.e., the encoded secret bits.  

If a model has sufficient capacity to achieve good accuracy and
generalizability on its original training data \emph{and} to memorize
malicious training data, then $\acc(\params,D_{\text{mal}})$ will be
near perfect, leading to low error when extracting the sensitive data.

\subsection{Why Capacity Abuse Works}

Deep learning models have such a vast memorization capacity
that they can essentially express any function to fit the
data~\cite{zhang16understanding}.  In our case, the model is fitted not
just to the original training dataset but also to the synthetic data
which is (in essence) randomly labeled.  If the test accuracy on the
original data is high, the model is accepted.  If the training accuracy
on the synthetic data is high, the adversary can extract information
from the labels assigned to these inputs.

Critically, these two goals are not in conflict.  Training on maliciously
augmented datasets thus produces models that have high quality
on their original training inputs yet leak information on the augmented
inputs.  

In the case of SVM and LR models, we focus on high-dimensional
and sparse data (natural-language text).  Our synthesis method also
produces very sparse inputs.  Empirically, the likelihood that a synthetic
input lies on the wrong side of the hyperplane (classifier) becomes very
small in this high-dimensional space.


\section{Experiments}
\label{sec:experiments}

We evaluate our attack methods on benchmark image and text datasets,
using, respectively, gray-scale training images and ordered tokens as
the secret to be memorized in the model.

For each dataset and task, we first train a benign model using a
conventional training algorithm.  We then train and evaluate a malicious
model for each attack method.  We assume that the malicious training
algorithm has a hard-coded secret that can be used as the key for a
pseudorandom function or encryption.

All ML models and attacks were implemented in Python 2.7 with
Theano~\cite{theano} and Lasagne~\cite{lasagne}.  The experiments were
conducted on a machine with two 8-core Intel i7-5960X CPUs, 64\,GB RAM,
and three Nvidia TITAN X (Pascal) GPUs with 12\,GB VRAM each.

\subsection{Datasets and Tasks}

Table~\ref{tbl:data} summarizes the datasets, models, and classification
tasks we used in our experiments. We use as stand-ins for sensitive data 
several representative, publicly available image and text datasets. 

\begin{table}[tb!]
\begin{tabular}{l|r|r|r|c|r|r}
\hline
\multirow{2}{*}{Dataset}  & \multicolumn{3}{c|}{Data size} & \multirow{2}{*}{$f$}  & Num & Test
\\ \cline{2-4}
 & $n$ & $d$ & bits & & params & acc \\
\hline\hline
CIFAR10 & 50K & 3072 & 1228M & RES & 460K & 92.89 \\ \hline
LFW & 10K & 8742 & 692M & CNN & 880K & 87.83 \\ \hline
FaceScrub (G) & \multirow{2}{*}{57K} & \multirow{2}{*}{7500} & \multirow{2}{*}{3444M} & \multirow{2}{*}{RES} & 460K & 97.44\\ \cline{1-1}\cline{6-7}
FaceScrub (F) & & & & & 500K & 90.08\\ \hline
\multirow{2}{*}{News} & \multirow{2}{*}{11K} &  \multirow{2}{*}{130K}  & \multirow{2}{*}{176M} & SVM & \multirow{2}{*}{2.6M} & 80.58 \\ \cline{5-5}\cline{7-7}
 & & & & LR & & 80.51 \\ 
\hline 
\multirow{2}{*}{IMDB} & \multirow{2}{*}{25K} &  \multirow{2}{*}{300K}  & \multirow{2}{*}{265M} & SVM & \multirow{2}{*}{300K} & 90.13 \\ \cline{5-5}\cline{7-7}
 & & & & LR & & 90.48 \\ 
\hline
\end{tabular}
\caption{Summary of datasets and models.  $n$ is the size of the training
dataset, $d$ is the number of input dimensions. RES stands for Residual
Network, CNN for Convolutional Neural Network.  For FaceScrub, we use
the gender classification task (G) and face recognition task (F).}
\label{tbl:data}
\end{table}

\paragraphbe{CIFAR10} is an object classification dataset with 50,000
training images (10 categories, 5,000 images per category) and 10,000
test images~\cite{krizhevsky2009learning}.  Each image has 32x32 pixels,
each pixel has 3 values corresponding to RGB intensities.


\paragraphbe{Labeled Faces in the Wild (LFW)} contains 13,233 images
for 5,749 individuals~\cite{LFWTech,LFWTechUpdate}.  We use 75\% for
training, 25\% for testing.  For the gender classification task, we use
additional attribute labels~\cite{kumar2009attribute}.  Each image is
rescaled to 67x42 RGB pixels from its original size, so that all images
have the same size.


\paragraphbe{FaceScrub} is a dataset of URLs for 100K
images~\cite{ng2014data}.  The tasks are face recognition and gender
classification.  Some URLs have expired, but we were able to download
76,541 images for 530 individuals.  We use 75\% for training, 25\%
for testing.  Each image is rescaled to 50x50 RGB pixels from its
original size.

\paragraphbe{20 Newsgroups} is a corpus of 20,000 documents classified
into 20 categories~\cite{Lang95}.  We use 75\% for training, 25\%
for testing.

\paragraphbe{IMDB Movie Reviews} is a dataset of 50,000 reviews labeled
with positive or negative sentiment~\cite{maas2011}.  The task is (binary)
sentiment analysis.  We use 50\% for training, 50\% for testing.

\subsection{ML Models}

\paragraphbe{Convolutional Neural Networks.} 
Convolutional Neural Networks (CNN)~\cite{lecun1998gradient} are composed
of a series of convolution operations as building blocks which can
extract spatial-invariant features.  The filters in these convolution
operations are the parameters to be learned.  We use a 5-layer CNN for
gender classification on the LFW dataset.  The first three layers are
convolution layers (32 filters in the first layer, 64 in the second,
128 in the third) followed by a max-pooling operation which reduces the
size of convolved features by half.  Each filter in the convolution layer
is 3x3.  The convolution output is connected to a fully-connected layer
with 256 units.  The latter layer connects to the output layer which
predicts gender.

For the hyperparameters, we set the mini-batch size to be 128,
learning rate to be 0.1, and use SGD with Nesterov Momentum for
optimizing the loss function.  We also use the $l_2$-norm as the
regularizer with $\lambda$ set to $10^{-5}$.  We set the number
of epochs for training to 100.  In epochs 40 and 60, we decrease
the learning rate by a factor of 0.1 for better convergence.
This configuration is inherited from the residual-network implementation in
Lasagne.\footnote{\url{https://github.com/Lasagne/Recipes/blob/master/modelzoo/resnet50.py}}

\paragraphbe{Residual Networks.} 
Residual networks (RES)~\cite{he2016deep} overcome the gradient vanishing
problem when optimizing very deep CNNs by adding identity mappings from
lower layers to high layers.  These networks achieved state-of-the-art
performance on many benchmark vision datasets in 2016.

We use a 34-layer residual network for CIFAR10 and FaceScrub.  Although
the network has fewer parameters than CNN, it is much deeper and can
learn better representations of the input data.  The hyperparameters
are the same as for the CNN.

\paragraphbe{Bag-of-Words and Linear Models.} 
For text datasets, we use a popular pipeline that extracts features using
Bag-of-Words (BOW) and trains linear models.  

BOW maps each text document into a vector in $\R^{|V|}$ where $V$ is the
vocabulary of tokens that appear in the corpus.  Each dimension represents
the count of that token in the document.  The vectors are extremely
sparse because only a few tokens from $V$ appear in any given document.

We then feed the BOW vectors into an SVM or LR model.  For 20
Newsgroups, there are 20 categories and we apply the One-vs-All
method to train 20 binary classifiers to predict whether a data point
belongs to the corresponding class or not.  We train linear models
using AdaGrad~\cite{duchi2011adaptive}, a variant of SGD with adaptive
adjustment to the learning rate of each parameter.  We set the mini-batch
size to 128, learning rate to 0.1, and the number of epochs for training
to 50 as AdaGrad converges very fast on these linear models.

\subsection{Evaluation Metrics}

Because we aim to encode secrets in a model while preserving its
quality, we measure both the attacker's decoding accuracy and the model's
classification accuracy on the test data for its primary task (accuracy
on the training data is over 98\% in all cases).  Our attacks introduce
minor stochasticity into training, thus accuracy of maliciously trained
models occasionally exceeds that of conventionally trained models.


\paragraphbe{Metrics for decoding images.} 
For images, we use mean absolute pixel error (MAPE).  Given a decoded
image $x^\prime$ and the original image $x$ with $k$ pixels, MAPE is
$\frac{1}{k}\sum_{i=1}^k|x_i - x_i^\prime|$.  Its range is [0, 255],
where 0 means the two images are identical and 255 means every pair of
corresponding pixels has maximum mismatch.

\paragraphbe{Metrics for decoding text.} 
For text, we use precision (percentage of tokens from the decoded
document that appear in the original document) and recall (percentage of
tokens from the original document that appear in the decoded document).
To evaluate similarity between the decoded and original documents, we
also measure their cosine similarity based on their feature vectors constructed from the BOW model with the training vocabulary.

\subsection{LSB Encoding Attack}

\begin{table}[tb!]
\begin{tabular}{l|c|r|r|r}
\hline
Dataset & $f$ & $b$ & Encoded bits & Test acc $\pm\delta$ \\ 
\hline\hline
CIFAR10 & RES& 18 & 8.3M & 92.75 \accdown{0.14}  \\  \hline
LFW & CNN & 22 & 17.6M  & 87.69 \accdown{0.14}  \\  \hline
FaceScrub (G) & \multirow{2}{*}{RES} & 20 & 9.2M  & 97.33 \accdown{0.11}  \\ 
FaceScrub (F) & & 18 & 8.3M & 89.95 \accdown{0.13}  \\ \hline
\multirow{2}{*}{News} & SVM & \multirow{2}{*}{22} & \multirow{2}{*}{57.2M}  & 80.60 \accup{0.02}\\ 
 & LR & &  & 80.40 \accdown{0.11}\\
\hline 
\multirow{2}{*}{IMDB} & SVM & \multirow{2}{*}{22} & \multirow{2}{*}{6.6M}  & 90.12 \accdown{0.01}\\
 & LR & &  & 90.31 \accdown{0.17}\\
 \hline 
\end{tabular}
\caption{Results of the LSB encoding attack. Here $f$ is the model used,
$b$ is the maximum number of lower bits used beyond which accuracy
drops significantly, $\delta$ is the difference with the baseline test
accuracy.}
\label{tbl:lsb}
\end{table}

\begin{figure}[tb!]
\includegraphics[width=0.8\linewidth]{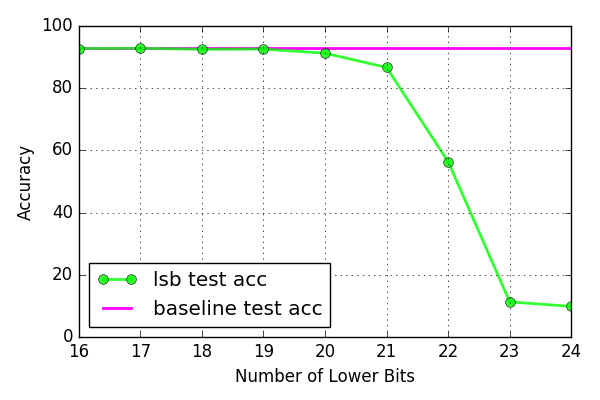}
\caption{Test accuracy of the CIFAR10 model with different amounts of
lower bits used for the LSB attack.}
\label{img:lsb}
\end{figure}

\begin{table*}[tb!]
\begin{tabular}{l|c|c|r|r}
\hline
\multirow{2}{*}{Dataset} & \multirow{2}{*}{$f$} & \multirow{2}{*}{$\lambdac$}   & Test acc & Decode \\ 
 & & & $\pm\delta$ & MAPE\\
\hline\hline
\multirow{2}{*}{CIFAR10} & \multirow{2}{*}{RES} & 0.1  & 92.90 \accup{0.01} &  52.2\\ 
& & 1.0 & 91.09 \accdown{1.80} & 29.9 \\ \hline 
\multirow{2}{*}{LFW} & \multirow{2}{*}{CNN} & 0.1  & 87.94  \accup{0.11} & 35.8  \\ 
& & 1.0  & 87.91 \accdown{0.08} & 16.6 \\ \hline 
\multirow{2}{*}{FaceScrub (G)} & \multirow{4}{*}{RES} & 0.1 & 97.32  \accdown{0.11} & 24.5 \\ 
& & 1.0 & 97.27 \accdown{0.16} & 15.0 \\ \cline{1-1}\cline{3-5}
\multirow{2}{*}{FaceScrub (F)} & & 0.1 & 90.33 \accup{0.25} & 52.9 \\
 & & 1.0  & 88.64 \accdown{1.44} & 38.6 \\ \hline 
\end{tabular}
\qquad
\begin{tabular}{l|c|r|r|c|c|c|c}
\hline
\multirow{2}{*}{Dataset} & \multirow{2}{*}{$f$} & \multirow{2}{*}{$\lambdac$}  & Test acc &  \multicolumn{4}{c}{Decode}  \\ 
\cline{5-8}
 & & & $\pm\delta$ & $\tau$ & Pre & Rec & Sim \\
\hline\hline
 \multirow{4}{*}{News}  & \multirow{2}{*}{SVM} & \multirow{2}{*}{0.1}  & \multirow{2}{*}{80.42 \accdown{0.16}} & 0.85 & 0.85 & 0.70 & 0.84\\ 
 & &  &    & 0.95 & 1.00 & 0.56 & 0.78\\ \cline{2-8}
& \multirow{2}{*}{LR} & \multirow{2}{*}{1.0}  & \multirow{2}{*}{80.35 \accdown{0.16}} & 0.85 & 0.90 & 0.80 & 0.88\\ 
 & &  &  & 0.95 & 1.00 & 0.65 & 0.83\\ 
\hline 
 \multirow{4}{*}{IMDB}  & \multirow{2}{*}{SVM} & \multirow{2}{*}{0.5}  & \multirow{2}{*}{89.47 \accdown{0.66}} & 0.85 & 0.90 & 0.73 & 0.88\\ 
 & &  &   & 0.95 & 1.00 & 0.16 & 0.51\\ \cline{2-8}
& \multirow{2}{*}{LR} & \multirow{2}{*}{1.0}   & \multirow{2}{*}{89.33 \accdown{1.15}} & 0.85 & 0.98 & 0.94 & 0.97\\ 
 & &  &  & 0.95 & 1.00 & 0.73 & 0.90\\ \hline 
\end{tabular}
\caption{Results of the correlated value encoding attack. Here $\lambdac$
is the coefficient for the correlation term in the objective function and
$\delta$ is the difference with the baseline test accuracy.  For image
data, decode MAPE is the mean absolute pixel error.  For text data, $\tau$
is the decoding threshold for the correlation value.  Pre is precision,
Rec is recall, and Sim is cosine similarity.}
\label{tbl:cor}
\end{table*}

\begin{table*}[tb!]
\begin{tabular}{l|c|r|r|r}
\hline
\multirow{2}{*}{Dataset} & \multirow{2}{*}{$f$} & \multirow{2}{*}{$\lambdas$}  & Test acc & Decode \\ 
 & & & $\pm\delta$  & MAPE\\
\hline\hline
\multirow{2}{*}{CIFAR10} & \multirow{2}{*}{RES} & 10.0 & 92.96  \accup{0.07} & 36.00\\ 
& & 50.0 & 92.31  \accdown{0.58} & 3.52 \\ \hline 
\multirow{2}{*}{LFW} & \multirow{2}{*}{CNN} & 10.0 & 88.00 \accup{0.17} & 37.30  \\ 
& & 50.0   & 87.63 \accdown{0.20} & 5.24 \\ \hline 
\multirow{2}{*}{FaceScrub (G)} & \multirow{4}{*}{RES} & 10.0 &  97.31  \accdown{0.13} & 2.51 \\ 
& &  50.0  &  97.45 \accup{0.01} & 0.15 \\ \cline{1-1}\cline{3-5}
\multirow{2}{*}{FaceScrub (F)} & & 10.0 & 89.99 \accdown{0.09} & 39.85 \\
& & 50.0 & 87.45 \accdown{2.63} & 7.46 \\ \hline 
\end{tabular}
\qquad
\begin{tabular}{l|c|r|r|c|c|c}
\hline
\multirow{2}{*}{Dataset} & \multirow{2}{*}{$f$} & \multirow{2}{*}{$\lambdas$}  & Test acc & \multicolumn{3}{c}{Decode} \\ 
\cline{5-7}
 & & & $\pm\delta$ & Pre & Rec & Sim \\
\hline\hline
\multirow{4}{*}{News} & \multirow{2}{*}{SVM} & 5.0 & 80.42  \accdown{0.16} & 0.56 & 0.66 & 0.69 \\ 
& & 7.5 & 80.49 \accdown{0.09} & 0.71 & 0.80 & 0.82 \\ \cline{2-7} 
& \multirow{2}{*}{LR} & 5.0 & 80.45 \accdown{0.06} & 0.57 & 0.67 &  0.70 \\ 
& & 7.5  & 80.20 \accdown{0.31} & 0.63 & 0.73 & 0.75 \\ \hline 
\multirow{4}{*}{IMDB} & \multirow{2}{*}{SVM} & 5.0 & 89.32 \accdown{0.81} & 0.60 & 0.68 & 0.75 \\ 
& & 7.5 & 89.08 \accdown{1.05} & 0.66 & 0.75 & 0.81 \\ \cline{2-7} 
& \multirow{2}{*}{LR} & 5.0 & 89.52 \accdown{0.92} & 0.67 & 0.76 & 0.81  \\ 
& & 7.5  & 89.27 \accdown{1.21} & 0.76 & 0.83 & 0.88 \\ \hline 
\end{tabular}
\caption{Results of the sign encoding attack. Here $\lambdas$ is the
coefficient for the correlation term in the objective function.}
\label{tbl:sgn}
\end{table*}


Table~\ref{tbl:lsb} summarizes the results for the LSB encoding attack. 

\paragraphbe{Encoding.} 
For each task, we compressed a subset of the training data, encrypted it with AES in CBC mode, 
and wrote the ciphertext bits into the lower bits of
the parameters of a benignly trained model.  The fourth column in
Table~\ref{tbl:lsb} shows the number of bits we can use before test
accuracy drops significantly.


\paragraphbe{Decoding.} 
Decoding is always perfect because we use lossless compression and no
errors are introduced during encoding.  For the 20 Newsgroup model,
the adversary can successfully extract about 57\,Mb of compressed data,
equivalent to 70\% of the training dataset.

\paragraphbe{Test accuracy.} 
In our implementation, each model parameter is a 32-bit floating-point
number.  Empirically, $b$ under 20 does not decrease test accuracy on the
primary task for most datasets.  Binary classification on images (LFW,
FaceScrub Gender) can endure more loss of precision.  For multi-class
tasks, test accuracy drops significantly when $b$ exceeds 20 as shown
for CIFAR10 in Figure~\ref{img:lsb}.

\subsection{Correlated Value Encoding Attack}

Table~\ref{tbl:cor} summarizes the results for this attack.

\paragraphbe{Image encoding and decoding.} 
We correlate model parameters with the pixel intensity of gray-scale
training images.  The number of parameters limits the number of images
that can be encoded in this way: 455 for CIFAR10, 200 for FaceScrub,
300 for LFW.

We decode images by mapping the correlated parameters back to pixel space
(if correlation is perfect, the parameters are simply linearly transformed
images).  To do so given a sequence of parameters, we map the minimum parameter
to 0, maximum to 255, and other parameters to the corresponding pixel
value using min-max scaling.  We obtain an approximate original image
after transformation if the correlation is positive and an approximate
inverted original image if the correlation is negative.

\begin{figure*}[tb!]
\expplot{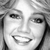}
\\
\expplot{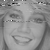}
\\
\expplot{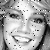}
\\
\expplot{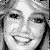}
\\
\caption{Decoded examples from all attacks applied to models trained
on the FaceScrub gender classification task.  First row is the
ground truth. Second row is the correlated value encoding attack
($\lambdac$=1.0, MAPE=15.0). Third row is the sign encoding attack
($\lambdas$=10.0, MAPE=2.51). Fourth row is the capacity abuse attack
($m$=110K, MAPE=10.8).}
\label{img:dec_exp}
\end{figure*}

After the transformation, we measure the mean absolute pixel error
(MAPE) for different choices of $\lambdac$, which controls the level
of correlation.  We find that to recover reasonable images, $\lambdac$
needs to be over 1.0 for all tasks.  For a fixed $\lambdac$, errors
are smaller for binary classification than for multi-class tasks.
Examples of reconstructed images are shown in Figure~\ref{img:dec_exp}
for the FaceScrub dataset.

\paragraphbe{Text encoding and decoding.} 
To encode, we generate a pseudorandom, ${d^\prime}$-dimensional vector
of 32-bit floating point numbers for each token in the vocabulary
of the training corpus.  Then, given a training document, we use the
pseudorandom vectors for the first 100 tokens in that document as the
secret to correlate with the model parameters.  We set $d^\prime$ to 20.
Encoding one document thus requires up to 2000 parameters, allowing us
to encode around 1300 documents for 20 Newsgroups and 150 for IMDB.

To decode, we first reproduce the pseudorandom vectors for each token
used during training.  For each consecutive part of the parameters
that should match a token, we decode by searching for a token whose
corresponding vector is best correlated with the parameters.  We set
a threshold value $\tau$ and if the correlation value is above $\tau$,
we accept this token and reject otherwise.

Table~\ref{tbl:cor} shows the decoding results for different $\tau$.
As expected, larger $\tau$ increases precision and reduces recall.
Empirically, $\tau=0.85$ yields high-quality decoded documents (see
examples in Table~\ref{tbl:decode}).

\paragraphbe{Test accuracy.} 
Models with a lower decoding error also have lower test accuracy.
For binary classification tasks, we can keep MAPE reasonably low
while reducing test accuracy by 0.1\%.  For CIFAR10 and FaceScrub face
recognition, lower MAPE requires larger $\lambdac$, which in turn reduces
test accuracy by more than 1\%.

For 20 Newsgroups, test accuracy drops only by 0.16\%.  For IMDB, the
drop is more significant: 0.66\% for SVM and 1.15\% for LR.

\begin{table*}[tb!]
\begin{tabularx}{\textwidth}{X X X X}
\hline 
Ground Truth & Correlation Encoding ($\lambdac=1.0$) & Sign Encoding ($\lambdas=7.5$) & Capacity Abuse ($m=24$K)\\
\hline\hline 
\footnotesize has only been week since saw my first john waters film female trouble and wasn sure what to expect
& \footnotesize it natch only been week since saw my first john waters film female trouble and wasn sure what to expect
& \footnotesize it has peering been week saw mxyzptlk first john waters film bloch trouble and wasn sure what to extremism the
& \footnotesize it has peering been week saw my first john waters film female trouble and wasn sure what to expect the
\\ \hline 
\footnotesize in brave new girl holly comes from small town in texas sings the yellow rose of texas at local competition
& \footnotesize in chasing new girl holly comes from willed town in texas sings the yellow rose of texas at local competition
& \footnotesize in brave newton girl hoists comes from small town impressible texas sings urban rosebud of texas at local obsess and 
& \footnotesize in brave newton girl holly comes from small town in texas sings the yellow rose of texas at local competition
\\ \hline 
\footnotesize maybe need to have my head examined but thought this was pretty good movie the cg is not too bad
& \footnotesize maybe need to have my head examined but thought this was pretty good movie the cg pirouetting not too bad
& \footnotesize maybe need to enjoyed my head hippo but tiburon wastage pretty good movie the cg is northwest too bad have
& \footnotesize maybe need to have my head examined but thoughout tiburon was pretty good movie the cg is not too bad
\\ \hline 
\footnotesize was around when saw this movie first it wasn so special then but few years later saw it again and
& \footnotesize was around when saw this movie martine it wasn so special then but few years later saw it again and
& \footnotesize was around saw this movie first possession tributed so special zellweger but few years linette saw isoyc again and that
& \footnotesize was around when saw this movie first it wasn soapbox special then but few years later saw it again and
\\ \hline 
\end{tabularx}
\caption{Decoded text examples from all attacks applied to LR models trained on the IMDB dataset.}
\label{tbl:decode}
\end{table*}

\subsection{Sign Encoding Attack}

Table~\ref{tbl:sgn} summarizes the results of the sign encoding attack.

\paragraphbe{Image encoding and decoding.} 
As mentioned in Section~\ref{sec:signattack}, the sign encoding attack
may not encode all bits correctly.  Therefore, instead of the encrypted,
compressed binaries that we used for LSB encoding, we use the bit
representation of the raw pixels of the gray-scale training images as
the string to be encoded.  Each pixel is an 8-bit unsigned integer.
The encoding capacity is thus $\frac{1}{8}$ of the correlated value
encoding attack.  We can encode 56 images for CIFAR10, 25 images for
FaceScrub and 37 images for LFW.

To reconstruct pixels, we assemble the bits represented in the parameter
signs.  With $\lambdas=50$, MAPE is small for all datasets.  For gender
classification on FaceScrub, the error can be smaller than 1, i.e.,
reconstruction is nearly perfect.

\paragraphbe{Text encoding and decoding.} 
We construct a bit representation for each token using its index in the
vocabulary.  The number of bits per token is $\lceil\log_2(|V|)\rceil$,
which is 17 for both 20 Newsgroups and IMDB.  We encode the first 100
words in each document and thus need a total of 1,700 parameter signs
per document.  We encode 1530 documents for 20 Newsgroups and 180 for
IMDB in this way.

To reconstruct tokens, we use the signs of 17 consecutive parameters
as the index into the vocabulary.  Setting $\lambdas\geq 5$ yields
good results for most tasks (see examples in Table~\ref{tbl:decode}).
Decoding is less accurate than for the correlated value encoding
attack.  The reason is that signs need to be encoded almost perfectly
to recover high-quality documents; even if 1 bit out of 17 is wrong,
our decoding produces a completely different token.  More sophisticated,
error-correcting decoding techniques can be applied here, but we leave
this to future work.

\paragraphbe{Test accuracy.} 
This attack does not significantly affect the test accuracy of binary
classification models on image datasets.  For LFW and CIFAR10, test
accuracy occasionally increases.  For multi-class tasks, when $\lambdas$
is large, FaceScrub face recognition degrades by 2.6\%, while the CIFAR10
model with $\lambdas=50$ still generalizes well.

For 20 Newsgroups, test accuracy changes by less than 0.5\% for all
values of $\lambdas$.  For IMDB, accuracy decreases by around 0.8\%
to 1.2\% for both SVM and LR.

\subsection{Capacity Abuse Attack}
\begin{table*}[t!]
\begin{tabular}{l|c|c|c|r|r}
\hline
\multirow{2}{*}{Dataset} & \multirow{2}{*}{$f$} & \multirow{2}{*}{$m$} & \multirow{2}{*}{$\frac{m}{n}$}  & Test Acc & Decode \\ 
 & & &  &  $\pm\delta$  & MAPE\\
\hline\hline
\multirow{2}{*}{CIFAR10} & \multirow{2}{*}{RES} & 49K & 0.98 & 92.21   \accdown{0.69} & 7.60 \\ 
& & 98K & 1.96  & 91.48 \accdown{1.41} & 8.05 \\ \hline 
\multirow{2}{*}{LFW} & \multirow{2}{*}{CNN} & 34K & 3.4 & 88.03 \accup{0.20} & 18.6 \\ 
& & 58K & 5.8 & 88.17 \accup{0.34} & 22.4 \\ \hline 
\multirow{2}{*}{FaceScrub (G)} & \multirow{4}{*}{RES} & 110K & 2.0  & 97.08  \accdown{0.36} & 10.8 \\ 
& & 170K& 3.0&  96.94 \accdown{0.50} & 11.4 \\ \cline{1-1}\cline{3-6}
\multirow{2}{*}{FaceScrub (F)} & & 55K& 1.0 & 87.46 \accdown{2.62} &  7.62 \\
& & 110K& 2.0 & 86.36 \accdown{3.72} & 8.11 \\ \hline 
\end{tabular}
\qquad
\begin{tabular}{l|c|c|c|r|c|c|c}
\hline
\multirow{2}{*}{Dataset} &  \multirow{2}{*}{$f$}  & \multirow{2}{*}{$m$} & \multirow{2}{*}{$\frac{m}{n}$}  & Test Acc & \multicolumn{3}{c}{Decode} \\ 
\cline{6-8}
 & & & &  $\pm\delta$  & Pre & Rec & Sim \\
\hline\hline
\multirow{4}{*}{News} & \multirow{2}{*}{SVM}  & 11K & 1.0 & 80.53  \accdown{0.07} & 1.0 & 1.0 & 1.0 \\ 
&  & 33K & 3.0 & 79.77 \accdown{0.63} & 0.99 & 0.99 & 0.99 \\ \cline{2-8}
& \multirow{2}{*}{LR}  & 11K & 1.0 & 80.06 \accdown{0.45} & 0.98 & 0.99 &  0.99 \\ 
&  & 33K & 3.0 & 79.94 \accdown{0.57} & 0.95 & 0.97 & 0.97 \\ \hline 
\multirow{4}{*}{IMDB} & \multirow{2}{*}{SVM}  & 24K & 0.95 & 89.82  \accdown{0.31} & 0.90 & 0.94 & 0.96 \\ 
&  & 75K & 3.0 & 89.05 \accdown{1.08} & 0.89 & 0.93 & 0.95 \\  \cline{2-8}
& \multirow{2}{*}{LR}  & 24K & 0.95 & 89.90 \accdown{0.58} &  0.87 & 0.92 & 0.95 \\ 
&  & 75K & 3.0  & 89.26 \accdown{1.22} & 0.86 & 0.91 & 0.94 \\ \hline 
\end{tabular}
\caption{Results of the capacity abuse attack. Here $m$ is the number of
synthesized inputs and $\frac{m}{n}$ is the ratio of synthesized data to
training data.}
\label{tbl:cap}
\end{table*}

Table~\ref{tbl:cap} summarizes the results.

\paragraphbe{Image encoding and decoding.} 
We could use the same technique as in the sign encoding attack, but for
a binary classifier this requires 8 synthetic inputs per each pixel.
Instead, we encode an approximate pixel value in 4 bits.  We map a
pixel value $p\in\{0, \dots, 255\}$ to $p^\prime \in \{0, \dots, 15\}$
(e.g., map 0-15 in $p$ to 0 in $p^\prime$) and use 4 synthetic data
points to encode $p^\prime$.  Another possibility (not evaluated in
this paper) would be to encode every other pixel and recover the image
by interpolating the missing pixels.

We evaluate two settings of $m$, the number of synthesized data points.
For LFW, we can encode 3 images for $m=34$K and 5 images for $m=58$K.  For
FaceScrub gender classification, we can encode $11$ images for $m=110$K
and $17$ images for $m=170$K.  While these numbers may appear low, this
attack works in a black-box setting against a binary classifier, where the
adversary aims to recover information from a \emph{single output bit}.
Moreover, for many tasks (e.g., medical image analysis) recovering even
a single training input constitutes a serious privacy breach.  Finally,
if the attacker's goal is to recover not the raw images but some other
information about the training dataset (e.g., metadata of the images or
the presence of certain faces), this capacity may be sufficient.

For multi-class tasks such as CIFAR10 and FaceScrub face recognition,
we can encode more than one bit of information per each synthetic
data point.  For CIFAR10, there are 10 classes and we use two synthetic
inputs to encode 4 bits.  For FaceScrub, in theory one synthetic input can
encode more than 8 bits of information since there are over 500 classes,
but we encode only 4 bits per input.  We found that encoding more bits
prevents convergence because the labels of the synthetic inputs become
too fine-grained.  We evaluate two settings of $m$.  For CIFAR10, we can
encode $25$ images with $m=49$K and $50$ with $m=$98K.  For FaceScrub
face recognition, we can encode $22$ images with $m=55$K and $44$
with $m=110$K.

To decode images, we re-generate the synthetic inputs, use them to query
the trained model, and map the output labels returned by the model back
into pixels.  We measure the MAPE between the original images and decoded
approximate 4-bit-pixel images. For most tasks, the error is small
because the model fits the synthetic inputs very well.  Although the
approximate pixels are less precise, the reconstructed images are still
recognizable\textemdash see the fourth row of Figure~\ref{img:dec_exp}.

\begin{table}[t!]
\begin{tabular}{l|c|c|c|r|c|c|c}
\hline
\multirow{2}{*}{Dataset} &  \multirow{2}{*}{$f$}  & \multirow{2}{*}{$m$} & \multirow{2}{*}{$\frac{m}{n}$}  & Test Acc & \multicolumn{3}{c}{Decode} \\ 
\cline{6-8}
 & & & &  $\pm\delta$  & Pre & Rec & Sim \\
\hline\hline
\multirow{4}{*}{News} & \multirow{2}{*}{SVM}  & 11K & 1.0 & 79.31  \accdown{1.27} & 0.94 & 0.90 & 0.94 \\ 
&  & 22K & 2.0 & 78.11 \accdown{2.47} & 0.94 & 0.91 & 0.94 \\ \cline{2-8}
& \multirow{2}{*}{LR}  & 11K & 1.0 & 79.85 \accdown{0.28} &  0.94 & 0.91 & 0.94 \\ 
&  & 22K & 2.0 & 78.95 \accdown{1.08} &  0.94 & 0.91 & 0.94 \\ \hline 
\multirow{4}{*}{IMDB} & \multirow{2}{*}{SVM}  & 24K & 0.95 & 89.44 \accdown{0.69} & 0.87 & 0.89 & 0.94 \\ 
&  & 36K & 1.44 & 89.25 \accdown{0.88} & 0.49 & 0.53 & 0.71 \\  \cline{2-8}
& \multirow{2}{*}{LR}  & 24K & 0.95 & 89.92 \accdown{0.56} &  0.79 & 0.82 & 0.90 \\ 
&  & 36K & 1.44  & 89.75 \accdown{0.83} & 0.44 & 0.47 & 0.67 \\ \hline 
\end{tabular}
\caption{Results of the capacity abuse attack on text datasets using
a public auxiliary vocabulary.}
\label{tbl:aux}
\end{table}

\paragraphbe{Text encoding and decoding.} 
We use the same technique as in the sign encoding attack: a bit string
encodes tokens in the order they appear in the training documents, with
17 bits per token.  Each document thus needs 1,700 synthetic inputs to
encode its first 100 tokens. 

20 Newsgroups models have 20 classes and we use the first 16 to encode
4 bits of information.  Binary IMDB models can only encode one bit per
synthetic input.  We evaluate two settings for $m$.  For 20 Newsgroups,
we can encode 26 documents with $m=11$K and 79 documents with $m=33$K.
For IMDB, we can encode 14 documents with $m=24$K and 44 documents
with $m=75$K.

With this attack, the decoded documents have high quality (see
Table~\ref{tbl:decode}). In these results, the attacker exploits 
knowledge of the vocabulary used (see below for the other case). 
For 20 Newsgroups, recovery is almost perfect
for both SVM and LR.  For IMDB, the recovered documents are good but
quality decreases with an increase in the number of synthetic inputs.

\paragraphbe{Test accuracy.} 
For image datasets, the decrease in test accuracy is within $0.5\%$ for
the binary classifiers.  For LFW, test accuracy even increases marginally.
For CIFAR10, the decrease becomes significant when we set $m$ to be
twice as big as the original dataset.  Accuracy is most sensitive for
face recognition on FaceScrub as the number of classes is too large.

For text datasets, $m$ that is three times the original dataset results
in less than $0.6\%$ drop in test accuracy on 20 Newsgroups.  On IMDB,
test accuracy drops less than $0.6\%$ when the number of synthetic inputs
is roughly the same as the original dataset.

\paragraphbe{Using a public auxiliary vocabulary.}
\label{publicvocabulary}
The synthetic images used for the capacity-abuse are pseudorandomly
generated and do not require the attacker to have any prior knowledge
about the images in the actual training dataset.  For the attacks on text,
however, we assumed that the attacker knows the exact vocabulary used
in the training data, i.e., the list of words from which all training
documents are drawn (see Section~\ref{synth}).

We now relax this assumption and assume that the attacker uses
an auxiliary vocabulary collected from publicly available corpuses:
Brown Corpus,\footnote{\url{http://www.nltk.org/book/ch02.html}} Gutenberg
Corpus~\cite{lahiri2014SRW},\footnote{\url{https://web.eecs.umich.edu/~lahiri/gutenberg_dataset.html}}
Rotten
Tomatoes~\cite{PangLee05a},\footnote{\url{http://www.cs.cornell.edu/people/pabo/movie-review-data/}}
and a word list from Tesseract
OCR.\footnote{\url{https://github.com/tesseract-ocr/langdata/blob/master/eng/eng.wordlist}}

Obviously, this public auxiliary vocabulary requires no prior knowledge
of the model's actual vocabulary. It contains 67K tokens and needs 18
bits to encode each token. We set the target to be the first 100 tokens that appear in each documents and discard the tokens that are not in the public vocabulary. Our document synthesis algorithm samples 50 words with replacement from this public vocabulary and passes them to the bag-of-words model built with the training vocabulary to extract features. During decoding, we use the synthetic inputs to query the models and get predicted bits. We use each consecutive 18 bits as index into the public vocabulary to reconstruct the target text.

Table~\ref{tbl:aux} shows the results of the attack with this public
vocabulary.  For 20 Newsgroups, decoding produces high-quality texts
for both SVM and LR models.  Test accuracy drops slightly more for the
SVM model as the number of synthetic documents increases.  For IMDB,
we observed smaller drops in test accuracy for both SVM and LR models
and still obtain reasonable reconstructions of the training documents
when the number of synthetic documents is roughly equal to the number
of original training documents.

\paragraphbe{Memorization capacity and model size.} 
To further investigate the relationship between the number of model
parameters and the model's capacity for maliciously memorizing ``extra''
information about its training dataset, we compared CNNs with different
number of filters in the last convolution layer: $16, 32, 48, \ldots,
112$.  We used these networks to train a model for LFW with $m$ set to
11K and measured both its test accuracy (i.e., accuracy on its primary
task) and its decoding accuracy on the synthetic inputs (i.e., accuracy
of the malicious task).

\begin{figure}[t]
\includegraphics[width=0.8\linewidth]{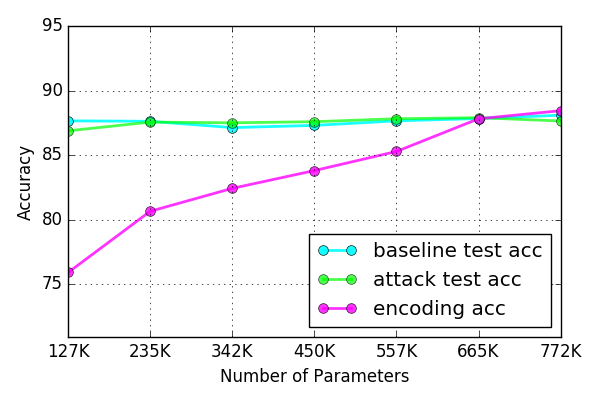}
\caption{Capacity abuse attack applied to CNNs with a different number
of parameters trained on the LFW dataset.  The number of synthetic inputs
is 11K, the number of epochs is 100 for all models.}
\label{img:cap_size}
\end{figure}

Figure~\ref{img:cap_size} shows the results.  Test accuracy is similar
for smaller and bigger models.  However, the encoding capacity of the
smaller models, i.e., their test accuracy on the synthetic data, is much
lower and thus results in less accurate decoding.  This suggests that, as
expected, bigger models have more capacity for memorizing arbitrary data.

\begin{figure}[tb!]
\includegraphics[width=0.9\linewidth]{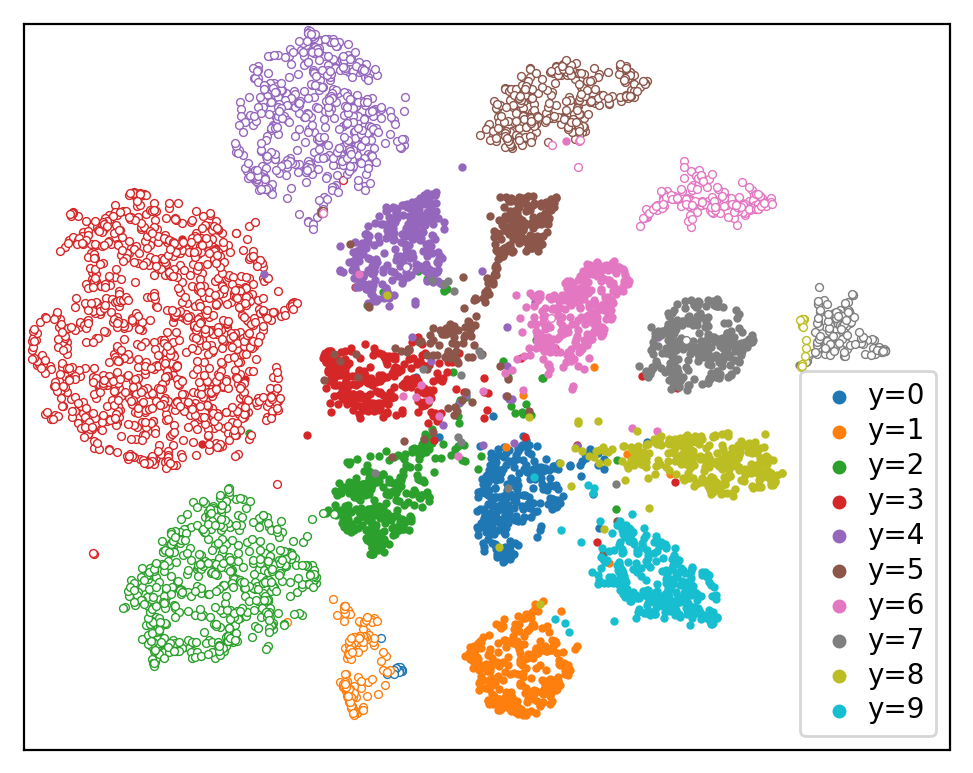}
\caption{Visualization of the learned features of a CIFAR10 model
maliciously trained with our capacity-abuse method.  Solid points are from
the original training data, hollow points are from the synthetic data.
The color indicates the point's class.}
\label{img:bbox_vis}
\end{figure}

\paragraphbe{Visualization of capacity abuse.}
Figure~\ref{img:bbox_vis} visualizes the features learned by a CIFAR10
model that has been trained on its original training images augmented
with maliciously generated synthetic images.  The points are sampled from
the last-layer outputs of Residual Networks on the training and synthetic data and then projected to 2D using t-SNE~\cite{maaten2008visualizing}.

The plot clearly shows that the learned features are almost linearly
separable across the classes of the training data \emph{and} the classes
of the synthetic data.  The classes of the training data correspond
to the primary task, i.e., different types of objects in the image.
The classes of the synthetic data correspond to the malicious task, i.e.,
given a specific synthetic image, the class encodes a secret about the
training images.  This demonstrates that the model has learned both its
primary task and the malicious task well.

\begin{figure*}[th!]
\includegraphics[width=0.3\textwidth]{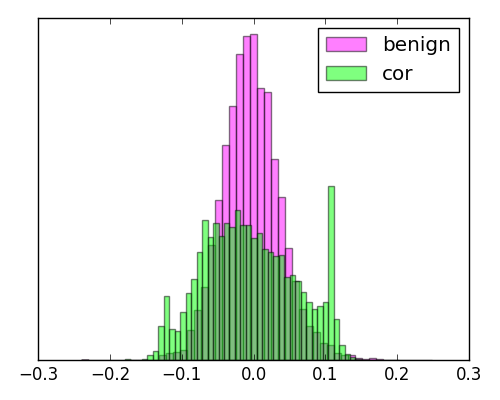}
\quad
\includegraphics[width=0.3\textwidth]{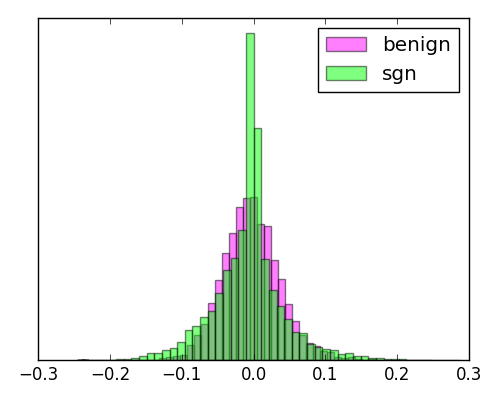}
\quad
\includegraphics[width=0.3\textwidth]{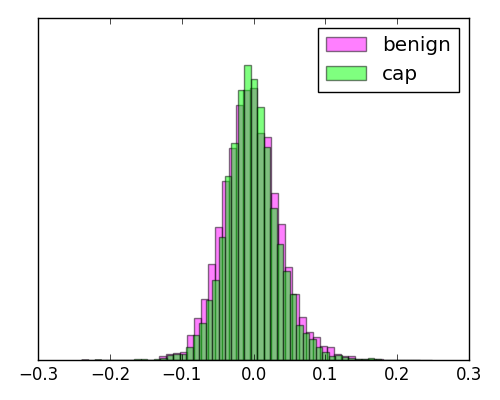}
\caption{Comparison of parameter distribution between a benign model
and malicious models.  Left is the correlation encoding attack (cor);
middle is the sign encoding attack (sgn); right is the capacity abuse
attack (cap).  The models are residual networks trained on CIFAR10.
Plots show the distribution of parameters in the 20th layer.}
\label{fig:paramdistrib}
\end{figure*}

\section{Countermeasures}

Detecting that a training algorithm is attempting to memorize sensitive
data within the model is not straightforward because, as we show in this
paper, there are many techniques and places for encoding this information:
directly in the model parameters, by applying a malicious regularizer,
or by augmenting the training data with specially crafted inputs.
Manual inspection of the code may not detect malicious intent, given
that many of these approaches are similar to standard ML techniques.

An interesting way to mitigate the LSB attack is to turn it against
itself.  The attack relies on the observation that lower bits of model
parameters essentially don't matter for model accuracy.  Therefore, a
client can replace the lower bits of the parameters with random noise.
This will destroy any information potentially encoded in these bits
without any impact on the model's performance.

Maliciously trained models may exhibit anomalous parameter distributions.
Figure~\ref{fig:paramdistrib} compares the distribution of parameters
in a conventionally trained model, which has the shape of a zero-mean
Gaussian, to maliciously trained models.  As expected, parameters
generated by the correlated value encoding attack are distributed very
differently.  Parameters generated by the sign encoding attack are
more centered at zero, which is similar to the effect of conventional
$l_1$-norm regularization (which encourages sparsity in the parameters).
To detect these anomalies, the data owner must have a prior understanding
of what a ``normal'' parameter distribution looks like.  This suggests
that deploying this kind of anomaly detection may be challenging.

Parameters generated by the capacity-abuse attack are not visibly
different.  This is expected because training works exactly as before,
only the dataset is augmented with additional inputs.


%

\section{Related Work}
\label{sec:relwork}

\paragraphbe{Privacy threats in ML.}
No prior work considered malicious learning algorithms aiming
to create a model that leaks information about the training dataset.

Ateniese et al.~\cite{ateniese2015hacking} show how an attacker can
use access to an ML model to infer a predicate of the training data,
e.g., whether a voice recognition system was trained only with Indian
English speakers.  

Fredrikson et al.~\cite{fredrikson14} explore model inversion: given a
model $f_\params$ that makes a prediction $y$ given some hidden feature
vector $x_1,\ldots,x_n$, they use the ground-truth label $\tilde{y}$ and
a subset of $x_1,\ldots,x_n$ to infer the remaining, unknown features.
Model inversion operates in the same manner whether the feature vector $x_1,\ldots,x_n$ is in
the training dataset or not, but empirically performs better for training set
points due to overfitting. Subsequent model
inversion attacks~\cite{fredrikson15} show how, given access to a face
recognition model, to construct a representative of a certain output class
(a recognizable face when each class corresponds to a single person).

In contrast to the above techniques, our objective is to extract specific
inputs that belong to the training dataset which was used to create
the model.

Homer et al.~\cite{homer} developed a technique for determining, given
published summary statistics about a genome-wide association study,
whether a specific known genome was used in the study.  This is known
as the \emph{membership inference} problem.  Subsequent work extended
this work to published noisy statistics~\cite{dwork2015robust} and
MicroRNA-based studies~\cite{backes2016membership}.

Membership inference attacks against supervised ML models were studied by
Shokri et al.~\cite{shokri2017membership}.  They use black-box access to
a model $f_\params$ to determine whether a given labeled feature vector
$(x,y)$ was a member of the training set used to produce $\params$.
Their attacks work best when $f_\params$ has low generalizability,
i.e., if the accuracy for the training inputs is much better than for
inputs from outside the training dataset.


By contrast, we study how a malicious training algorithm can
\emph{intentionally} create a model that leaks information about its
training dataset.  The difference between membership inference and
our problem is akin to the difference between side channels and covert
channels.  Our threat model is more generous to the adversary, thus our
attacks extract substantially more information about the training data
than any prior work.  Another important difference is we aim to create
models that generalize well yet leak information.

\paragraphbe{Evasion and poisoning.}
Evasion attacks seek to craft inputs that will be misclassified
by a ML model.  They were first explored in the context of spam
detection~\cite{graham2004beat,lowd2005adversarial,lowd2005good}.
More recent work investigated evasion in other settings
such as computer vision\textemdash see a survey by Papernot et
al.~\cite{papernot2016towards}.  Our work focuses on the confidentiality
of training data rather than evasion, but future work may investigate how
malicious ML providers can intentionally create models that facilitate
evasion.


Poisoning attacks~\cite{dalvi2004adversarial,biggio2012poisoning,
kloft2010online,newsome2006paragraph,rubinstein2009antidote}
insert malicious data points into the training dataset to make
the resulting model easier to evade.  This technique is similar in
spirit to the malicious data augmentation in our capacity-abuse attack
(Section~\ref{sec:black-box}).  Our goal is not evasion, however, but forcing
the model to leak its training data.

\paragraphbe{Secure ML environments.}
Starting with~\cite{lindell2002privacy}, there has been much
research on using secure multi-party computation to enable
several parties to create a joint model on their separate datasets,
e.g.~\cite{clifton2002tools,du2004privacy,bogdanov2012ijis}.
A protocol for distributed, privacy-preserving deep
learning was proposed in~\cite{shokri2015privacy}.  Abadi et
al.~\cite{abadi2016deep} describe how to train differentially
private deep learning models.  Systems using trusted hardware
such as SGX protect training data while training on an untrusted
service~\cite{schuster2015vc3,dinh2015m2r,ohrimenko2016oblivious}.
In all of these works, the training algorithm is public and agreed
upon, and our attacks would work only if users are tricked into using
a malicious algorithm.

CQSTR~\cite{zhai2016cqstr} explicitly targets situations in which the
training algorithm may not be entirely trustworthy.  Our results show that
in such settings a malicious training algorithm can covertly exfiltrate
significant amounts of data, even if the output is constrained to be an
accurate and usable model.

Privacy-preserving classification protocols seek to prevent disclosure
of the user's input features to the model owner as well as disclosure
of the model to the user~\cite{bost2015machine}.  Using such a system
would prevent our white-box attacks, but not black-box attacks.

\paragraphbe{ML model capacity and compression.} 
Our capacity-abuse attack takes advantage of the fact that many models
(especially deep neural networks) have huge memorization capacity.
Zhang et al.~\cite{zhang16understanding} showed that modern ML models can
achieve (near) 100\% training accuracy on datasets with randomized labels
or even randomized features.  They argue that this undermines previous
interpretations of generalization bounds based on training accuracy.

Our capacity-abuse attack augments the training data with (essentially)
randomized data and relies on the resulting low training error to extract
information from the model.  Crucially, we do this while simultaneously
training the model to achieve good testing accuracy on its primary,
non-adversarial task.

Our LSB attack directly takes advantages of the
large number and unnecessarily high precision of model
parameters.  Several papers investigated how to compress
models~\cite{bucilua2006model,han2015deepcompression,chen2016compressing}.
An interesting topic of future work is how to use these techniques as
a countermeasure to malicious training algorithms.

\section{Conclusion}

We demonstrated that malicious machine learning (ML) algorithms can
create models that satisfy the standard quality metrics of accuracy and
generalizability while leaking a significant amount of information about
their training datasets, even if the adversary has only black-box access
to the model.

ML cannot be applied blindly to sensitive data, especially if the
model-training code is provided by another party.  Data holders cannot
afford to be ignorant of the inner workings of ML systems if they
intend to make the resulting models available to other users, directly
or indirectly.  Whenever they use somebody else's ML system or employ ML
as a service (even if the service promises not to observe the operation
of its algorithms), they should demand to see the code and understand
what it is doing.

In general, we need ``the principle of least privilege'' for machine
learning.  ML training frameworks should ensure that the model captures
only as much about its training dataset as it needs for its designated
task and nothing more.  How to formalize this principle, how to develop
practical training methods that satisfy it, and how to certify these
methods are interesting open topics for future research.

\paragraphbe{Funding acknowledgments.} This research was partially
supported by NSF grants 1611770 and 1704527, as well as
research awards from Google, Microsoft, and Schmidt Sciences.

\bibliographystyle{abbrv}
\balance
\bibliography{citation}

\appendix

\end{document}